# Removing motion artifacts from mechanomyographic signals: an innovative filtering method applied to human movement analysis

Matthieu Correa[1], Nicolas Vignais[2], Isabelle A Siegler[3] and Maxime Projetti[1]

*1) Moten Technologies, France*
*2) Univ. Rennes, M2S, INRIA, F-35000 Rennes, France*
*3) Laboratoire CIAMS, Université Paris-Saclay, France*

*Abstract—* **Mechanomyography (MMG) is a promising tool for measuring muscle activity in the field but its sensitivity to motion artifacts limits its application. In this study, we proposed an adaptative filtering method for MMG accelerometers based on the complete ensemble empirical mode decomposition, with adaptative noise and spectral fuzzy entropy, to isolate motions artefacts from the MMG signal in dynamic conditions. We compared our method with the traditional band-pass filtering technique, demonstrating better results concerning motion recomposition for deltoid and erector spinae muscles ($R^2$ = 0.907 and 0.842). Thus, this innovative method allows the filtering of motion artifacts dynamically in the 5-20 Hz bandwidth, which is not achievable with traditional method. However, the interpretation of accelerometric MMG signals from the trunk and lower-limb muscles during walking or running should be approached with great caution as impact-related accelerations are still present, though their exact quantity still needs to be quantified.**

## II. INTRODUCTION

Measuring muscle activity in the field may permit to prevent work related musculoskeletal disorders (WMSD). Metrics related to muscle activity may be used to identify difficult working conditions and/or muscle fatigue, which is a known risk factor of WMSD. For this measurement, electromyography (EMG), which is defined as the measurement of muscle electrical currents, is the reference method. However, EMG requires skin preparation (e.g. shaving and disinfection) and it may be sensitive to electrical interferences like a change in impedance linked to sweating [1]. These limitations do not make EMG suitable for ecological situations such as an occupational environment, or a sport field in which sweating may occur.

The mechanical counterpart of EMG is mechanomyography (MMG), which is defined as the measurement of the low-frequency lateral oscillations of active skeletal muscle fibers [2]. MMG is generated by three principal components: i) a gross lateral movement of the muscle at the initiation of a contraction that is generated by non-simultaneous activation of muscle fibers, ii) smaller subsequent lateral oscillations occurring at the resonant frequency of the muscle, and iii) dimensional changes of the active muscle fibers [3]. MMG does not require skin preparation and is not sensitive to sweating nor electrical interferences, making it a promising alternative to EMG in the field.

However, accelerometers, which are the most common sensors for MMG measurements, are highly sensitive to motion artifacts. Motion artifacts are referred to as the active limbs motion and impacts [4]. To remove such artifacts (primarily the active limbs motion), previous studies historically used band-pass filters with band-pass values generally set at 5 and 100 Hz [5]. This filtering method has been extensively used in a majority of MMG studies regardless of contraction type. However it has been shown that in dynamic conditions, motion artifacts related to active limbs motion persist beyond 5 Hz (up to 15 Hz) [6], thus, altering the MMG signal despite filtering. To overcome this limitation during dynamic conditions, new filtering methods have been developed in recent studies [7], [8]. One of the most promising approaches is based on the Empirical Mode Decomposition (EMD). EMD is an iterative algorithm that decomposes a time series in subsequent Intrinsic Modes Functions (IMF) adaptatively without a priori knowledge, like Fourier based methods [9]. EMD general equation of a time series signal $S(t)$ is defined as follows:

$$S(t) = \sum_{i=1}^{n} IMF(t) + r(t) \quad (1)$$

where $r(t)$ is the decomposition residual and $n$ the total number of IMF. The main limitation of the EMD algorithm is related to the mode mixing problem, which is the presence of a same frequency component in different modes, removing the physical meaning of IMFs. To overcome mode mixing, some approaches suggested to add white noise into the original signal, thus permitting to efficiently decompose both synthetic and physiological signals compared to the original EMD [10]. This process has been implemented into the improved complete ensemble EMD with adaptative noise (iCEEMDAN) [11]. For the rest of this report, iCEEMDAN will be referred to as CEEMDAN. The general equation of the CEEMDAN decomposition of a time series signal $S(t)$ is defined as follows:

$$S(t) = \sum_{i=1}^{n} \overline{IMF_i(t)} + r_i(t) \quad (2)$$

where $n$ is the total number of IMF, $r$ the residue and $\overline{IMF_i}$ the true IMF obtained by averaging all samples of the ensemble at index $i$. An ensemble of size $N$ consists of the realization of $N$ signals with distinct additions of white noise.

The main drawback associated with the CEEMDAN method is related to the fact that a large ensemble size is critical to remove the white noise into the final decomposition, as it statistically cancels out with averaging. Depending on the complexity of the signal and the size of the ensemble, the CEEMDAN signal decomposition is very time consuming. For example, a signal of 5000 samples is decomposed in 30 seconds with an ensemble size of 500 [12]. This limitation makes CEEMDAN hardly usable in practical applications with extensive acquisition durations. To overcome this limitation, a GPU implementation of the CEEMDAN has been recently published, achieving up to 260 × speed-up compared to the reference MATLAB CPU method [12].

A recent study demonstrated how CEEMDAN can be used to filter motion artifacts from MMG signal during isometric conditions [7]. However, to the best of our knowledge, no studies were carried out in dynamic conditions to assess the effectiveness of a CEEMDAN based filtering method for MMG signals in these conditions.

In this context, the objective of this study was to propose a new filtering method based on CEEMDAN and compare it to traditional band-pass filtering. To this aim, muscle activity of three muscles were obtained with MMG accelerometers during a repeated load-lifting task.

III. MATERIAL AND METHODS

A. Participants

Twenty-one healthy subjects (9 females and 12 males) participated in the study. Written informed consent was given prior to the experiment. This study was approved by the French National research ethics committee of sciences and techniques of physical and sporting activities (CERSTAPS IRB00012476-2023-08-02-227).

B. Sensors

MMG signals from the biceps brachii (BB), lateral deltoid (DEL) and erector spinae (ES) muscles were recorded using accelerometric sensors (Moten Technologies, Puteaux, France) at 1000 Hz. Each sensor was placed on the muscle belly according to literature recommendations [13], [14]. A force sensor (2715-ISO, Sensy, Jumet, Belgium) was used to measure external force at the wrist with a sampling frequency of 2000 Hz during the maximum voluntary contractions (MVC). The MMG sensors were placed only on the dominant arm. Since only the Z component of the tri-axial accelerometer was used to measure MMG signal, its orientation did not affect MMG measurements.

C. Experimental design

After a standardized articular and muscular warm-up, the maximal voluntary contraction (MVC) measurements of the muscles studied were carried out. For the BB muscle, the participant had the elbow flexed at 90° against the body with the forearm in a neutral position and the force sensor placed at wrist level. For the DEL muscle, the arm was extended forward parallel to the ground with the force sensor on top of the wrist, facing downward. For the ES muscles, the participant was lying on their stomach with both hands close to the ears and was asked to push against the force sensor placed between their scapulas by extending the trunk. Each MVC lasted three seconds and was repeated three times with a 2-minute recovery period in between.

The weight carried during the load-lifting task corresponded to an intensity of 50% MVC of the ES muscle. The subject was asked to lift the load from a mark on the ground, carrying it and dropping it on another mark positioned 10 steps further. The subject was asked to repeat this task 24 times during a 4-minute cycle. A metronome was used to ensure a constant rate of 6 repetitions per minute throughout the experiment. Only the walking phase was analyzed in this study and was determined by computing the inclination angle of the trunk in the frontal plane from the acceleration data of the sensor placed on the ES muscle [15]. The portion of signal where the inclination angle of the trunk ranged between -10 and 10° was selected as walking.

D. CEEMDAN based filtering Method

The method proposed in this study, named $f_{ceemdan}$, consisted of: (i) decomposing the raw acceleration signal (see Figure 1) into IMFs with CEEMDAN, (ii) computing the power spectral density (PSD) of each IMF and its fuzzy entropy (FuzzEn), and (iii) recomposing the MMG signal by selecting adjacent IMFs with the IMF with the highest spectral FuzzEn value to maximize the quantity of MMG information (see Figure 2). Thus, the recomposed motion consisted of summing all higher order IMFs.

This method was compared to the reference MMG filtering method, consisting of a 4th order Butterworth band-pass filter between 5 and 100 Hz (named $f_{band}$) [5]. For the $f_{band}$, motion artifacts recomposition corresponded to the subtraction of the raw acceleration signal with the filtered MMG signal, following previous methodology [8].

E. Features extraction

To assess the performance of the $f_{ceemdan}$ compared to $f_{band}$, we computed the coefficient of determination ($R^2$) between motion artifacts recomposition methods and acceleration obtained from a reference motion sensor integrated in the MMG sensor (Z component of an accelerometer present in an Inertial Measurement Unit (IMU)). Moreover, to assess the quantity of motion artifacts filtered, we computed the absolute difference in PSD between each filtered MMG signal, and the raw acceleration ($\Delta_{psd}$). We divided the $\Delta_{psd}$ into subsequent frequency bandwidths (0-2.5; 2.5-5; 5-7.5; 7.5-10; 10-15; 15-20; 20-25: 25-30 Hz) to analyze the effect of the filtering methods across bandwidths, following previous methodology [8]

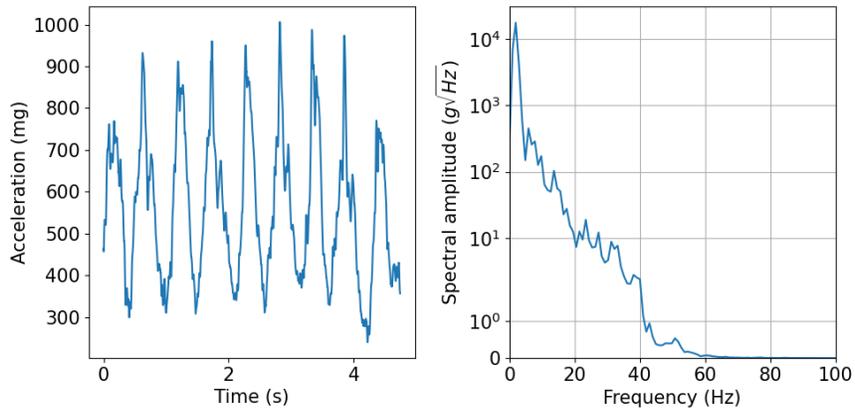

*Figure 1: Example of a raw acceleration signal from the Z component of the sensor placed on the BB muscle.*

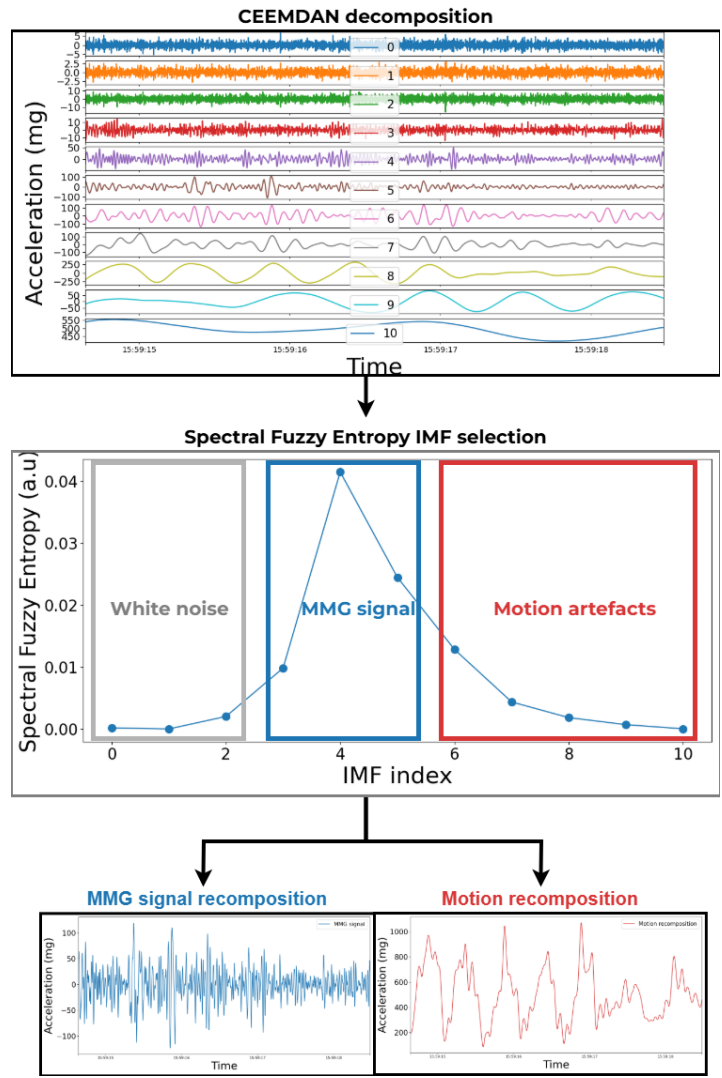

*Figure 2: Flowchart of the $f_{ceemdan}$ filtering method from a representative raw Z acceleration component from the BB muscle.*

*F. Statistical analysis*

In order to compare the motion artifacts recomposition scores (R²) of the two filtering methods for each muscle, a two-factor ANOVA (*method × muscle*) was conducted.

Besides, in order to compare the $\Delta_{psd}$ values between filtering methods and for each frequency band, a two-factor repeated measures ANOVA (*method × frequency band*) was conducted independently for each muscle. When necessary, post-hoc analysis was performed using Tukey's HSD. A significance level of α = 0.05 was used for statistical tests. Data are presented as mean ± standard deviation.

## IV. RESULTS

There was a significant effect of the filtering method and the muscle studied on the motion recomposition scores (p < 0.0001). Post-hoc analysis highlighted significantly higher motion artifacts recomposition scores for $f_{ceemdan}$ compared to $f_{band}$ for the DEL and ES muscles (p < 0.01 and p < 0.0001 respectively) (see Figure 3).

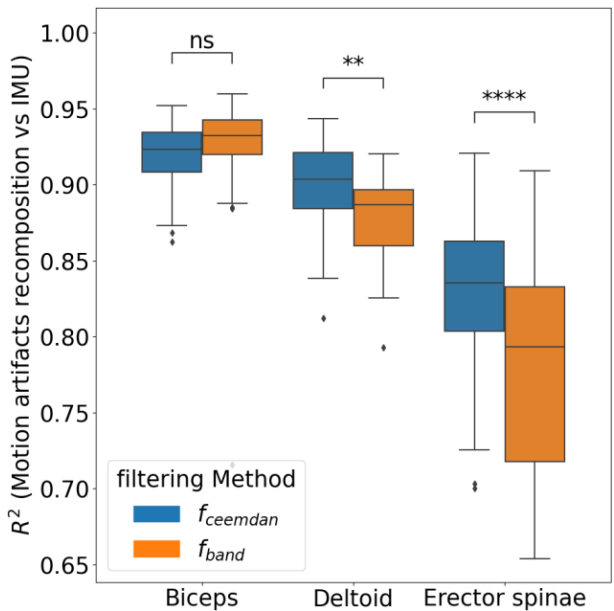

*Figure 3: Motion artifacts recomposition scores comparison (R² against the IMU reference) between filtering methods by muscle studied (**: p < 0.01; ****: p <0.0001).*

For all muscles, there was a significant effect of the frequency band and the filtering method on the $\Delta_{psd}$ variance (p < 0.0001). Post-hoc analysis highlighted that $\Delta_{psd}$ was significantly higher for $f_{ceemdan}$ in the frequency band 7.5 – 10 Hz for the BB muscle, 5 – 10 Hz for the DEL and 5 – 20 Hz or the ES muscle (see bottom chart from Figure 4).

Moreover, as shown in the top charts of Figure 4, the amplitude of the MMG signal obtained with $f_{ceemdan}$ is reduced (-34%, -28% and -49% for the BB, DEL and ES muscles respectively) compared to $f_{band}$, respectively. For $f_{ceemdan}$, the mean power frequency (MPF) of the filtered signal was 16 Hz for the BB, 14 Hz for DEL and 18 Hz for the ES muscles. For $f_{band}$, the MPF was 11 Hz for BB, 11 Hz for DEL and 12 Hz for the ES muscles. As a side note, it is interesting to point out that for both filtering methods, the majority of the signals spectral energy is distributed below 60 Hz, which is a well-known characteristic of MMG signals [2]

## V. DISCUSSION

The objective of the study was to propose a new filtering method based on CEEMDAN and fuzzy entropy and compare it to traditional band-pass filtering in dynamic conditions. To this aim, the muscle activity of three muscles (BB, DEL and ES muscles) were measured with MMG accelerometers during a repeated load-lifting task at constant speed and intensity.

Altogether, the results obtained in this study show that $f_{ceemdan}$ is more effective at filtering motions artifacts in dynamic conditions compared to $f_{band}$: $f_{ceemdan}$ motion recomposition scores are significantly higher for DEL and ES muscles, meaning that more motion artifacts are removed from the filtered MMG signal. This result is due to the filtering of motion artifacts beyond 5 Hz, as shown by the $\Delta_{psd}$ analysis with $f_{ceemdan}$. For the BB muscle, the lack of significant difference of motion recomposition scores between methods may be due to the lower local presence of motion artifacts in the first place, meaning that either filtering method may be equally effective in these conditions for this specific muscle. This result is also supported by the $\Delta_{psd}$ analysis where only the 7.5 – 10 Hz frequency bandwidth was subject to difference between methods.

A typical MMG signal obtained in isokinetic condition should have a MPF around 16 Hz [2], [3], [15]. Such properties are obtained by the mean of the $f_{ceemdan}$ method in contrast with $f_{band}$, where MPF values are close to 11 Hz (see Figure ). This result also corroborates the higher presence of motions artifacts (which have a lower frequency and greater amplitude than MMG [6]), in the filtered signal by $f_{band}$.

However, it was found that for the ES muscles, significantly lower recomposition scores were obtained compared to the motion artifacts reference compared to the BB and DEL muscles (- 9 points). This result highlights the fact that, despite a more effective filtering obtained by $f_{ceemdan}$ compared to $f_{band}$, motion artifacts are still present. It was shown in the literature that motion accelerations, corresponding to impacts, measured during walking at T12/L5 share a common frequency band with MMG between 10 and 30 Hz [4]. This result can be extended to lower limbs, which undergo higher levels of acceleration. Thus, the interpretation of accelerometric MMG signals in these areas when walking or running should be taken with great care.

To the best of our knowledge, no study has been conducted on the quantification of motion artifacts from accelerometric MMG signals during dynamic activities or the feasibility of unmixing motion artifacts and MMG signals beyond 5 Hz in such conditions. The current study established the relative effectiveness of the proposed method in comparison to the reference band-pass filtering method. However, as highlighted by ES muscles, residual motion artifacts remaining in the filtered MMG signal still need to be quantified.

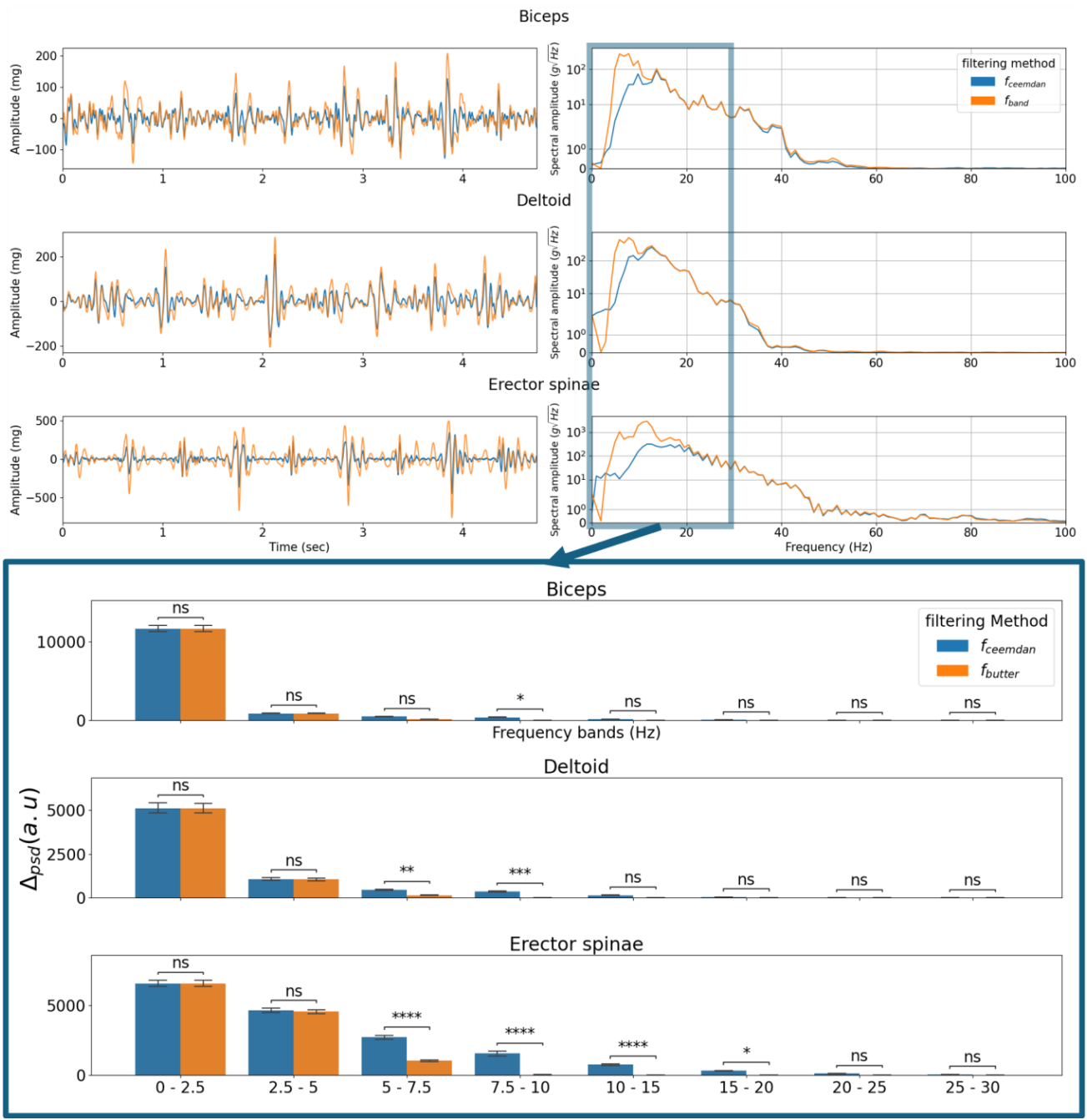

*Figure 4: Top charts: Example of filtered MMG signals in time (left) and frequency (right) domains for each muscle. PSDs are shown on a log scale because of the order of magnitude difference of spectral amplitudes between filtering methods. Bottom chart: comparisons of $\Delta_{psd}$ (absolute PSD difference between the filtered and the raw acceleration signal) between filtering methods and by frequency band (\*: $p < 0.05$; \*\*: $p < 0.01$; \*\*\* $p < 0.001$; \*\*\*\*: $p < 0.0001$*

## VI. Conclusion

The method proposed in this study showed its effectiveness in filtering motion artifacts during dynamic conditions, compared to reference method. This contribution opens the feasibility of using accelerometer MMG during functional activities in the field with promising applications in ergonomics, sports, rehabilitation or in clinical studies. Despite filtering, precautions should be taken when investigating MMG signals of trunk and lower-limb muscles when walking or running, as motion artifacts residues related to shocks and impact remain in the filtered signal. Future investigations should be carried out on this matter to better understand the properties of such artifacts onto the MMG signal and to assess the feasibility of unmixing motion artifacts and the MMG signal beyond 5 Hz.